\documentclass[12pt]{article}
\usepackage{graphicx}

\def\be{\begin{equation}}
\def\ee{\end{equation}}
\def\bea{\begin{eqnarray}}
\def\eea{\end{eqnarray}}
\def\bean{\begin{eqnarray*}}
\def\eean{\end{eqnarray*}}
\def\bary{\begin{array}}
\def\eary{\end{array}}

\def\bit{\begin{itemize}}
\def\eit{\end{itemize}}
\def\b{{\cal B}}

\def\vud{V_{ud}}
\def\vus{V_{us}}

\def\vcd{V_{cd}}
\def\vcs{V_{cs}}

\def\ub{{\bar u}}

\def\db{{\bar d}}
\def\sb{{\bar s}}

\begin{document}
\rightline{ANL-HEP-PR-01-098}
\rightline{EFI-01-48}
\rightline{hep-ph/0110394}
\rightline{October 2001}
\bigskip
\centerline{\bf FINAL-STATE PHASES IN DOUBLY-CABIBBO-SUPPRESSED}
\centerline{\bf CHARMED MESON NONLEPTONIC DECAYS}
\bigskip

\centerline {Cheng-Wei Chiang$\,^{a,b}$~\footnote{chengwei@hep.uchicago.edu}
and
Jonathan L. Rosner$\,^a$~\footnote{rosner@hep.uchicago.edu}}
\vspace{0.5cm}
\centerline{\it $^a$ Enrico Fermi Institute and Department of Physics}
\centerline{\it University of Chicago, 5640 S. Ellis Avenue, Chicago, IL 60637}
\vspace{0.2cm}
\centerline{\it $^b$ HEP Division, Argonne National Laboratory}
\centerline{\it 9700 S. Cass Avenue, Argonne, IL 60439}

\begin{quote}
Cabibbo-favored nonleptonic charmed particle decays exhibit large final-state
phase differences in $\bar K \pi$ and $\bar K^* \pi$ but not $\bar K \rho$
channels.  It is of interest to know the corresponding pattern of final-state
phases in doubly-Cabibbo-suppressed decays, governed by the $c \to d u \bar s$
subprocess.  An experimental program is outlined for determining such phases
via measurements of rates for $D \to K^* \pi$ and $K (\rho, \omega,\phi)$
channels, and determination of interference between bands in Dalitz plots.
Such a program is feasible at planned high-intensity sources of charmed
particles.
\end{quote}

\section{Introduction}

The observation of direct CP violation in decays of particles containing
heavy $(c,b)$ quarks requires two or more channels differing in both strong
and weak phases.  Whereas the weak phases can be anticipated within the
Standard Model based on the Cabibbo-Kobayashi-Maskawa matrix, the strong
phases must in general be extracted from experiment.  This is particularly
so in the case of charmed particle decays, where phases in some channels
have been shown to be large.  (For particles containing $b$ quarks, schemes
for calculating such phases have been proposed recently \cite{BBNS,KLS}.)

In Cabibbo-favored decays of charmed particles, governed by the subprocess $c
\to s u \bar d$, the pattern of final-state phases differs from channel to
channel.  In the decays $D \to \bar K \pi$ and $D \to \bar K^* \pi$, the
final states with isospins $I = 1/2$ and $I = 3/2$ have relative phases
close to $90^\circ$, while in $D \to \bar K \rho$, the $I = 1/2$ and $I = 3/2$
final states have relative phases close to zero.  This behavior has been
traced using an SU(3) flavor analysis \cite{Rosner:1999xd} to a sign flip in
the contribution of one of the amplitudes contributing to the $\bar K \rho$
processes in comparison with its contribution to the other two.

The corresponding final-state phases for doubly-Cabibbo-suppressed charmed
particle decays, governed by the subprocess $c \to d u \bar s$, are
of interest for several reasons.  First, they are needed whenever one wishes to
study CP asymmetries in such decays.  Such asymmetries are not expected in the
Standard Model, but the low rate for such processes makes them especially
sensitive in their CP asymmetries to non-standard contributions.  Second, the
question of whether final-state phases are the same in CP-conjugate states
such as $K^+ \pi^-$ and $K^- \pi^+$ \cite{CC,BP,FNP,GRU} is of current
interest in interpreting $D^0$--$\bar D^0$ mixing results.  Proposals for
shedding light on this question include using the correlations between $D^0$
and $\bar D^0$ at the $\psi(3770)$ \cite{GGR}, and assuming relations among
phase shifts in different $K^* \pi$ channels with the same isospin \cite{GP}.

It is easy to determine relative final-state phases in Cabibbo-favored
$D$ decays since there are three charge states (such as $D^0 \to K^- \pi^+$,
$D^0 \to \bar K^0 \pi^0$, and $D^+ \to \bar K^0 \pi^+$) and only two
independent amplitudes.  The amplitudes for the three processes thus form a
triangle in the complex plane as a result of the definite isospin of the $c \to
s u \bar d$ subprocess:  $\Delta I = \Delta I_3 = 1$.  We shall refer to such
decays as ``right-sign.''  In contrast, the subprocess $c \to d u \bar s$
governing doubly-Cabibbo-suppressed decays, which we shall call ``wrong-sign,''
has $\Delta I_3 = 0$
and either $\Delta I = 0$ or $\Delta I = 1$.  There are four charge states
(e.g., $D^0 \to K^+ \pi^-$, $D^0 \to K^0 \pi^0$, $D^+ \to K^+ \pi^0$, and
$D^+ \to K^0 \pi^+$) and three isospin amplitudes (two with $I = 1/2$ and
one with $I = 3/2$), so that the amplitudes form a quadrangle.  Without
additional assumptions or information, one cannot learn relative phases.

The right-sign amplitude triangle for two final-state pseudoscalar mesons
is related by a U-spin transformation \cite{Usp} $(d \leftrightarrow s)$ to
a corresponding triangle involving the two wrong-sign $D^0$ decays (to $K^+
\pi^-$ and $K^0 \pi^0$) and the decay $D_s \to K^0 K^+$ \cite{GRU}.  However,
the final states involving $K^0$ cannot be distinguished from the
much-more-copious right-sign final states involving $\bar K^0$.  If one
replaces a $K^0$ by a $K^{*0}$, one can learn its flavor by its decay to
$K^+ \pi^-$.  However, in the case of $D$ decays to a vector meson and a
pseudoscalar meson, the U-spin transformation turns out not to give a
useful relation because of the lack of symmetry under interchange of the two
final particles.  One can estimate final-state phases for the wrong-sign $D \to
K \pi$ decays with the help of information about direct-channel resonances
and form factors \cite{GRU}.

Using the wrong-sign decays $D \to K^* \pi$, for which one can determine the
flavor of the $K^*$ for all four charge states, Golowich and Pakvasa \cite{GP}
obtained a constraint sufficient to specify relative phases of amplitudes
(given measurements of all four rates) by assuming that the final-state phases
in the two $I = 1/2$ $K^* \pi$ amplitudes are equal.  Since this assumption is
risky for a highly inelastic channel such as $K^* \pi$ at the mass of the $D$,
we seek an alternative method which employs only experimental data.  We have
found such a method which relies upon interference of $K^*$ bands in the
$K^+ \pi^- \pi^0$ Dalitz plot.  In the course of this study, we find that
all the relative phases of wrong-sign $D$ decay amplitudes with one
pseudoscalar meson $P$ and one vector meson $V$ in the final state can be
specified using just $K \pi \pi$ and $K K \bar K$ final states.  These
predictions can then be checked in cases where a $\pi^0$ is replaced by an
$\eta$ or $\eta'$.

We begin in Section II with a decomposition of amplitudes for $D \to P P$ and
$D \to P V$ final states.  We point out relations among these in Section III,
and discuss experimental prospects for testing them in Section IV.
Section V concludes.

\section{Amplitude Decompositions}

We can categorize decay amplitudes according to the topology of Feynman
diagrams \cite{GHLR}: (1) a color-favored tree amplitude $T$, (2) a
color-suppressed tree amplitude $C$, (3) an exchange amplitude $E$, and (4) an
annihilation amplitude $A$.  $E$ only contributes to $D^0$ decays, and $A$ only
to Cabibbo-favored $D_s^+$ decays and Cabibbo-suppressed $D^+$ decays.  The
Cabibbo-favored non-leptonic two-body decays are governed by the subprocess $c
\to s u \db$ involving the weak coupling $\vcs^* \vud$, while the
doubly-Cabibbo-suppressed ones are governed by the subprocess $c \to d u \sb$
involving the weak coupling $\vcd^* \vus$.  We use notation introduced in
Ref.\ \cite{DGR98} for $PV$ decays in which a subscript denotes the meson ($P$
or $V$) containing the spectator quark. 

We can decompose the decay amplitudes both in terms of their topological
characters and in terms of isospin structure.  We use the
following quark content and phase conventions \cite{GHLR}:
\begin{itemize}
\item{
{\it Charmed mesons}: $D^0=-c\ub$, $D^+=c\db$, $D_s^+=c\sb$;}
\item{
{\it Pseudoscalar mesons}: $\pi^+=u\db$, $\pi^0=(d\db-u\ub)/\sqrt{2}$,
 $\pi^-=-d\ub$, $K^+=u\sb$, $K^0=d\sb$, ${\bar K}^0=s\db$, $K^-=-s\ub$,
 $\eta=(s\sb-u\ub-d\db)/\sqrt{3}$, $\eta^{\prime}=(u\ub+d\db+2s\sb)/\sqrt{6}$;}
\item{
{\it Vector mesons}: $\rho^+=u\db$, $\rho^0=(d\db-u\ub)/\sqrt{2}$,
 $\rho^-=-d\ub$, $\omega=(u\ub+d\db)/\sqrt{2}$, $K^{*+}=u\sb$, $K^{*0}=d\sb$,
 ${\bar K}^{*0}=s\db$, $K^{*-}=-s\ub$, $\phi=s\sb$.}
\end{itemize}
The wrong-sign (WS) $D$ decays are listed in Tables \ref{tab:WSPP} and
\ref{tab:WSVP}, where $SU(3)$ flavor symmetry is assumed.  We distinguish the
amplitudes obtained through $I=1$ and $I=0$ currents by superscripts 1 and 0
on the amplitudes $A_{1/2}$ and $B_{1/2}$.  We list the isospin decompositions
only for $K \pi$ and $K^* \pi$ modes.  It is the amplitudes $B^1_{1/2}$ and
$B^0_{1/2}$ which were assumed to have the same strong phases in Ref.\
\cite{GP}.  As mentioned, we make no such assumption.
For some of the other decays we list simplified expressions
which arise from assuming relations between different $E$ or $A$ amplitudes.
As in Ref.\ \cite{Rosner:1999xd}, we omit contributions of flavor topologies
in which $\eta$ and $\eta'$ exchange no quark lines with the rest of the
diagram, and couple through their SU(3)-singlet components.
This assumption, which goes beyond a purely SU(3)-based analysis, appeared to
give a self-consistent description in the case of most right-sign decays with
the exception of $D_s^+ \to \rho^+ \eta'$.  We shall see that it can be
tested in the case of WS decays, since the individual $T$, $C$, $E$, and $A$
amplitudes can be predicted independently of modes involving $\eta$ and
$\eta'$.

\section{Amplitude Relations}

{\footnotesize
\begin{table}
\caption{Amplitudes for WS decay modes of charmed mesons to two pseudoscalar
 mesons.
\label{tab:WSPP}}
\begin{center}
\begin{tabular}{lcc} \hline\hline
Mode & $A_{\rm topology}$ & $A_{\rm isospin}$ \\
\hline
$D^0 \to K^+\pi^-$ & $T+E$ & $\frac13 \left( A_{3/2}-A^1_{1/2} \right)-
\frac{1}{\sqrt{3}} A^0_{1/2}$ \\
$D^0 \to K^0\pi^0$ & $\frac{1}{\sqrt{2}}(C-E)$ & $\frac{\sqrt{2}}{3}A_{3/2}+
\frac{1}{3\sqrt{2}}A^1_{1/2}+\frac{1}{\sqrt{6}}A^0_{1/2} $ \\
$D^0 \to K^0\eta$ & $\frac{1}{\sqrt{3}}C$ &  \\
$D^0 \to K^0\eta^{\prime}$ & $-\frac{1}{\sqrt{6}}(C+3E)$ & \\
\hline
$D^+ \to K^0\pi^+$ & $C+A$ & $\frac13 \left( A_{3/2}-A^1_{1/2} \right)+
\frac{1}{\sqrt{3}}A^0_{1/2}$ \\
$D^+ \to K^+\pi^0$ & $\frac{1}{\sqrt{2}}(T-A)$ & $\frac{\sqrt{2}}{3}A_{3/2}+
\frac{1}{3\sqrt{2}}A^1_{1/2}-\frac{1}{\sqrt{6}}A^0_{1/2} $ \\
$D^+ \to K^+\eta$ & $-\frac{1}{\sqrt{3}}T$ &  \\
$D^+ \to K^+\eta^{\prime}$ & $\frac{1}{\sqrt{6}}(T+3A)$ & \\
\hline
$D_s^+ \to K^+K^0$ & $T+C$ & \\
\hline\hline
\end{tabular}
\end{center}
\end{table}
}

{\footnotesize
\begin{table}
\caption{Amplitudes for WS decay modes of charmed mesons to one
vector meson and one pseudoscalar meson.
\label{tab:WSVP}}
\begin{center}
\begin{tabular}{lll} \hline\hline
Mode & $A_{\rm topology}$ & $A_{\rm isospin}$ \\ \hline
$D^0 \to K^{*+}\,\pi^-$ & $T_P+E_V$ & 
$\frac13 \left( B_{3/2}-B^1_{1/2} \right)-\frac{1}{\sqrt{3}} B^0_{1/2}$ \\
$D^0 \to K^{*0}\,\pi^0$ & $\frac{1}{\sqrt{2}} \left( C_P-E_V \right)$ &
$\frac{\sqrt{2}}{3}B_{3/2}+\frac{1}{3\sqrt{2}}B^1_{1/2}+\frac{1}{\sqrt{6}}
B^0_{1/2}$ \\
$D^+ \to K^{*0}\,\pi^+$ & $C_P+A_V$ &
$\frac13 \left( B_{3/2}-B^1_{1/2} \right)+\frac{1}{\sqrt{3}}B^0_{1/2}$ \\
$D^+ \to K^{*+}\,\pi^0$ & $\frac{1}{\sqrt{2}} \left( T_P-A_V \right)$ &
$\frac{\sqrt{2}}{3}B_{3/2}+\frac{1}{3\sqrt{2}}B^1_{1/2}-\frac{1}{\sqrt{6}}
B^0_{1/2}$ \\
\hline
Mode & $A_{\rm topology}$ & $A_{\rm simplified}$ \\ \hline
$D^0 \to \phi\,K^0$ & $-E_V$ & \\
$D^0 \to \rho^-\,K^+$ & $T_V+E_P$ & $= T_V-E_V$ \\
$D^0 \to \rho^0\,K^0$ & 
$\frac{1}{\sqrt{2}}(C_V-E_P)$ & $= \frac{1}{\sqrt{2}}(C_V+E_V)$ \\
$D^0 \to \omega\,K^0$ & 
$-\frac{1}{\sqrt{2}}(C_V+E_P)$ & $= -\frac{1}{\sqrt{2}}(C_V-E_V)$ \\
$D^0 \to K^{*0}\,\eta$ & 
$\frac{1}{\sqrt{3}}(C_P-E_P+E_V)$ & $= \frac{1}{\sqrt{3}}(C_P+2E_V)$ \\
$D^0 \to K^{*0}\,\eta^{\prime}$ & 
$-\frac{1}{\sqrt{6}}(C_P+2E_P+E_V)$ & $= -\frac{1}{\sqrt{6}}(C_P-E_V)$ \\
\hline
$D^+ \to \phi\,K^+$ & $A_V$ & \\
$D^+ \to \rho^+\,K^0$ & $C_V+A_P$ & $= C_V-A_V$ \\
$D^+ \to \rho^0\,K^+$ & 
$\frac{1}{\sqrt{2}}(T_V-A_P)$ & $= \frac{1}{\sqrt{2}}(T_V+A_V)$ \\
$D^+ \to \omega\,K^+$ & 
$\frac{1}{\sqrt{2}}(T_V+A_P)$ & $= \frac{1}{\sqrt{2}}(T_V-A_V)$ \\
$D^+ \to K^{*+}\,\eta$ & 
$-\frac{1}{\sqrt{3}}(T_P-A_P+A_V)$ & $= -\frac{1}{\sqrt{3}}(T_P+2A_V)$ \\
$D^+ \to K^{*+}\,\eta^{\prime}$ & 
$\frac{1}{\sqrt{6}}(T_P+2A_P+A_V)$ & $= \frac{1}{\sqrt{6}}(T_P-A_V)$ \\
\hline
$D_s^+ \to K^{*+}\,K^0$ & $T_P+C_V$ & \\
$D_s^+ \to K^{*0}\,K^+$ & $T_V+C_P$ & \\
\hline\hline
\end{tabular}
\end{center}
\end{table}
}

The RS $D \to \bar K^* \pi$ decays give the sum rule
\be
\label{eqn:RS-DK*pi}
A(D^0 \to K^{*-}\pi^+) + \sqrt{2}A(D^0 \to {\bar K}^{*0}\pi^0)
- A(D^+ \to {\bar K}^{*0}\pi^+) = 0,
\ee
which forms a triangle in the amplitude complex plane.  This triangle, and
corresponding ones for $D \to \bar K \pi$ and $D \to \bar K \rho$, have been
used to obtain relative phases between the unique $I = 1/2$ and $I = 3/2$
amplitudes contributing to each set of processes \cite{Rosner:1999xd,Suz98}.

The sum rules for WS $D \to PP$ decays \cite{HJLPC},
\bea
3 \sqrt{2}A(K^+\pi^0) + 4\sqrt{3}A(K^+\eta) + \sqrt{6}A(K^+\eta^{\prime})
 &=& 0, \\
3 \sqrt{2}A(K^0\pi^0) - 4\sqrt{3}A(K^0\eta) - \sqrt{6}A(K^0\eta^{\prime}) &=&0,
\eea
allow one to form triangles.  
In terms of amplitudes of different topologies, these are, respectively,
\bea
3(T-A) - 4T + (T+3A) &=& 0, \\
3(C-E) - 4C + (C+3E) &=& 0.
\eea
The sum rules
\bea
&A(K^+\pi^-) + \sqrt{2}A(K^0\pi^0) 
= A(K^0\pi^+) + \sqrt{2}A(K^+\pi^0)& \nonumber \\
&= \sqrt{3}[A(K^0\eta)-A(K^+\eta)]
= A(K^+K^0)&
\label{eqn:wstrisum}
\eea
give triangles all sharing one side.
This can be seen from the decomposed amplitudes
\be
(T+E) + (C-E) = (C+A) + (T-A) = T+C.
\ee
We also find from these WS $D \to PP$ modes the following relations:
\bea
|T|^2 &=& 3|A(K^+\eta)|^2, \\
|C|^2 &=& 3|A(K^0\eta)|^2, \\
|A|^2 &=& \frac12 \left[ |A(K^+\pi^0)|^2 + |A(K^+\eta^{\prime})|^2 \right] -
 |A(K^+\eta)|^2, \\
|E|^2 &=& \frac12 \left[ |A(K^0\pi^0)|^2 + |A(K^0\eta^{\prime})|^2 \right] -
 |A(K^0\eta)|^2, \\
\cos\delta_{TC} &=& \frac{1}{2|T||C|}
\left[ |A(K^+K^0)|^2 - 3|A(K^+\eta)|^2 - 3|A(K^0\eta)|^2 \right], \\
\cos\delta_{TA} &=& \frac{1}{2|T||A|}
\left[ 2|A(K^+\eta)|^2 + \frac12|A(K^+\eta^{\prime})|^2 -
 \frac32|A(K^+\pi^0)|^2 \right], \\
\cos\delta_{CE} &=& \frac{1}{2|C||E|}
\left[ 2|A(K^0\eta)|^2 + \frac12|A(K^0\eta^{\prime})|^2 -
 \frac32|A(K^0\pi^0)|^2 \right], \\
\cos\delta_{TE} &=& \frac{1}{2|T||E|}
\left\{ |A(K^+\pi^-)|^2 - 3|A(K^+\eta)|^2 \right. \nonumber \\
&& \qquad \left. - \frac12 \left[ |A(K^0\pi^0)|^2 + |A(K^0\eta^{\prime})|^2
 \right] + |A(K^0\eta)|^2 \right\}, \\
\cos\delta_{CA} &=& \frac{1}{2|C||A|}
\left\{ |A(K^0\pi^+)|^2 - 3|A(K^0\eta)|^2 \right. \nonumber \\
&& \qquad \left. - \frac12 \left[ |A(K^+\pi^0)|^2 + |A(K^+\eta^{\prime})|^2
 \right] + |A(K^+\eta)|^2 \right\}.
\eea
Therefore, knowing the absolute value of the decay amplitudes one could
completely determine the above triangles.  However, all decays involving a
$K^0$ will be overwhelmed by Cabibbo-favored decays involving a $\bar K^0$,
with no way to distinguish between them since one detects only a $K_S$.  Thus
in practice one is able to determine only $|T|$, $|A|$, and $\delta_{TA}$,
which is still a useful piece of information relevant to final-state
interactions.  We shall discuss the prospects for this determination in
Section IV.

\begin{figure}
\centerline{\includegraphics[height=3in]{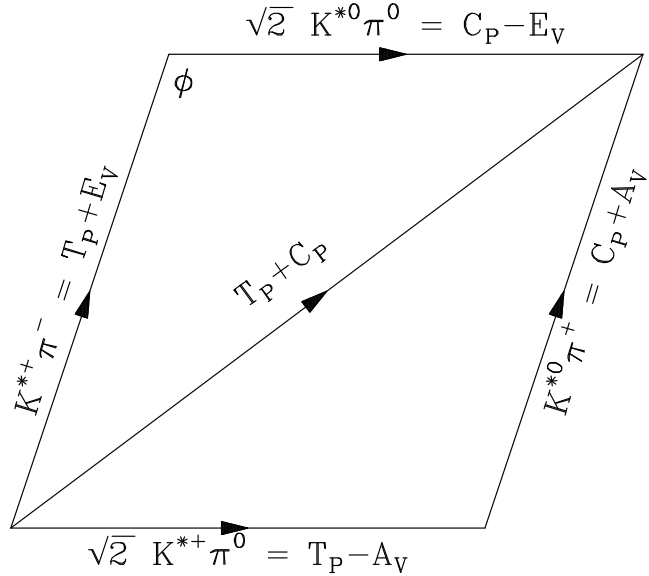}}
\caption{Quadrangle illustrating amplitude relations for $D \to K^* \pi$
decays.  The other diagonal (not shown) corresponds to the combination
$E_V + A_V$.
\label{fig:quad}}
\end{figure}

The WS $D \to K^* \pi$ decays give the sum rule
\bea
&A(K^{*+}\pi^-) + \sqrt{2}A(K^{*0}\pi^0)
= A(K^{*0}\pi^+) + \sqrt{2}A(K^{*+}\pi^0)&
\nonumber \\
&= (T_P + E_V) + (C_P - E_V) = (C_P + A_V) + (T_P - A_V) = T_P + C_P,&
\label{eqn:quadsum}
\eea
which forms a quadrangle in the complex plane, as shown in Fig.\
\ref{fig:quad}.

Knowing the lengths of the four sides in a quadrangle does not fix the shape;
one still needs information about relative angles among the sides.  In
principle such information could be obtained from other sum rules involving any two of the decay modes related to the sides of the quadrangle
in which we are interested.  However, these were searched for in Ref.\
\cite{GRU}, and no such triangle sum rule exists for these WS decays.

Fortunately, one can use interference between the two $K^*$ bands on the Dalitz
plot for $D^0 \to K^+ \pi^- \pi^0$, a final state recently reported by the CLEO
Collaboration \cite{Brandenburg:2001ze}, to measure the relative phase $\phi$
between the amplitudes for $D^0 \to K^{*+} \pi^-$ and $D^0 \to K^{*0} \pi^0$.
This method is analogous to the use of the decay $D^0 \to K_S \pi^+ \pi^-$
in which the interference between $K^{*+} \pi^-$ and $K^{*-} \pi^+$ bands
provides direct information on the relative strong phase difference between
the two channels \cite{DMA,Palano}.  Once the angle $\phi$ in Fig.\
\ref{fig:quad} is
specified, the shape of the quadrangle is fixed up to a folding about the
diagonal.  However, this is still not sufficient to specify each individual
amplitude $T_P$, $C_P$, $E_V$, or $A_V$.

One way to help resolve the above ambiguity is to compare the WS quadrangle
with the RS triangle [Eq.\ (\ref{eqn:RS-DK*pi})].  Denote the relative phase
between $D^0 \to K^{*-} \pi^+$ and $D^0 \to K^{*+} \pi^-$ by $\theta_0$, that
between $D^+ \to {\bar K}^{*0} \pi^+$ and $D^+ \to K^{*+} \pi^0$ by $\theta_+$,
and that between $D^0 \to K^{*-} \pi^+$ and $D^+ \to {\bar K}^{*0} \pi^+$ by
$\psi$.  $\theta_0$ can be obtained by analyzing the $K^{*+}$ and $K^{*-}$
bands in the Dalitz plot of the final state $D^0 \to K_S \pi^+ \pi^-$;
$\theta_+$ can be similarly measured from the Dalitz plot of $D^+ \to K_S \pi^+
\pi^0$.  With $\psi$ given by the RS triangle, the relative phase between $D^0
\to K^{*+} \pi^-$ and $D^+ \to K^{*+} \pi^0$ is then $\psi \pm |\theta_0| \pm
|\theta_+|$.  Therefore, except for singular cases, the angle between the left
and bottom sides of the quadrangle in Fig.\ \ref{fig:quad} can be determined.

One also makes further progress by assuming \cite{Rosner:1999xd} that (1)
$A_P=-A_V$
and/or (2) $E_P=-E_V$.  These assumptions are valid if these amplitudes involve
an intermediate quark-antiquark state \cite{HJLeta}.

\begin{figure}
\centerline{\includegraphics[height=3in]{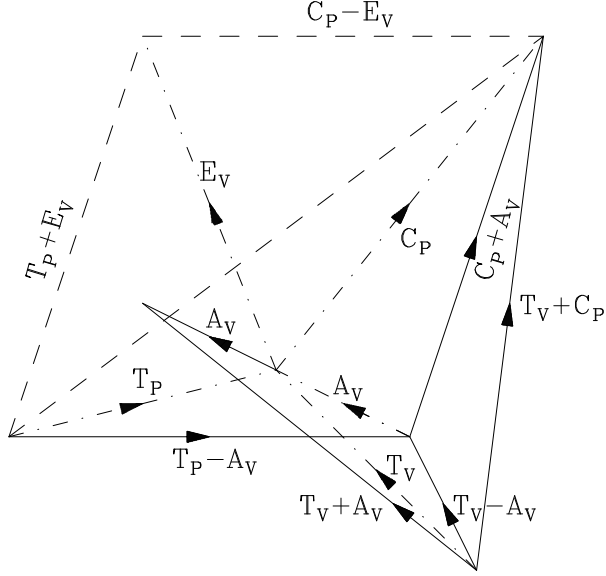}}
\caption{Amplitude triangles illustrating amplitude relations 
between $D^+ \to K^* \pi$ decays and other $D^+$ or $D_s^+$ decays.  The
dotdashed lines represent the individual amplitudes.
\label{fig:quadtri}}
\end{figure}

If only $A_P=-A_V$ is imposed, several of the expressions for
$D^+$ decays are simplified.  We find $A(K^{*+}\pi^0) = \sqrt{3}A(K^{*+}
\eta^{\prime})$ and the following sum rules:
\bea
A(K^{*0}K^+) - \sqrt{2}A(\omega K^+) - A(K^{*0}\pi^+) &=& 0,
\label{eqn:xsumDp1} \\
\sqrt{2}A(\rho^0 K^+) - \sqrt{2}A(\omega K^+) - 2A(\phi K^+) &=& 0,
\label{eqn:xsumDp2} \\
\sqrt{3}A(K^{*+} \eta) + \sqrt{2}A(K^{*+}\pi^0) + 3A(\phi K^+) &=& 0.
\label{eqn:xsumDp3}
\eea
In terms of amplitudes, these read, respectively,
\bea
(T_V + C_P) - (T_V - A_V) - (C_P + A_V) & = & 0, \\
(T_V + A_V) - (T_V - A_V) - 2 A_V       & = & 0, \\
-(T_P + 2 A_V) + (T_P - A_V) + 3 A_V    & = & 0.
\eea
The first two of these are illustrated in Fig.\ \ref{fig:quadtri}. Measurement
of the corresponding rates for $D_s \to K^{*0}K^+$ and $D^+ \to (\rho^0,\omega,
\phi) K^+$ along with the four $D \to K^* \pi$ rates and the relative phase of
$D^0 \to K^{*+} \pi^-$ and $D^0 \to K^{*0} \pi^0$ mentioned earlier can
specify the individual amplitudes up to the discrete ambiguity associated
with reflection about the dashed diagonal of the quadrangle.  This ambiguity
affects only the phase and magnitude of $E_V$ with respect to the other
amplitudes.  Since we have not used Eq.\ (\ref{eqn:xsumDp3}) in this
construction, we obtain a prediction for the amplitude $A(K^{*+} \eta)$.
The residual ambiguity can be removed if one assumes a certain magnitude
hierarchy among $T$, $C$ and $E$.

Under the assumption $A_P = - A_V$ we also find from the WS $D^+ \to VP$ modes
the following relations:
\bea
|A_V|^2 &=& |A(\phi K^+)|^2, \\
|T_V|^2 &=& |A(\rho^0 K^+)|^2 + |A(\omega K^+)|^2 - |A(\phi K^+)|^2, \\
|T_P|^2 &=& 4|A(K^{*+}\eta^{\prime})|^2 + |A(K^{*+}\eta)|^2 - 2|A(\phi K^+)|^2, \\
\cos\delta_{T_V A_V} &=& \frac{1}{2|T_V||A_V|}
\left[ |A(\rho^0 K^+)|^2 - |A(\omega K^+)|^2 \right], \\
\cos\delta_{T_P A_V} &=& \frac{1}{2|T_P||A_V|}
\left[ |A(K^{*+}\eta)|^2 - 2|A(K^{*+}\eta^{\prime})|^2 - |A(\phi K^+)|^2
 \right].
\eea
As in the WS $D^+ \to PP$ decays, we can learn both the magnitudes and the
relative phases of the $T$ and $A$ amplitudes directly from decay rates
involving observable final states.
 
If now $E_P = - E_V$ is assumed, some of the expressions in $D^0$ decays are
simplified.  One finds $A(K^{*0}\pi^0) = -\sqrt{3}A(K^{*0}\eta^{\prime})$ and
the following sum rules:
\bea
A(K^{*+}\pi^-)-\sqrt{2}A(\omega K^0)-A(K^{*+}K^0)&=&0,
\label{eqn:xsumD01} \\
\sqrt{2}A(\rho^0K^0)+\sqrt{2}A(\omega K^0)+2A(\phi K^0)&=&0,
\label{eqn:xsumD02} \\
\sqrt{3}A(K^{*0}\eta)-\sqrt{2}A(K^{*0}\pi^0)+3A(\phi K^0)&=&0.
\label{eqn:xsumD03}
\eea
These have the following form in terms of amplitudes:
\bea
(T_P + E_V) + (C_V - E_V) - (T_P + C_V) & = & 0, \\
(C_V + E_V) - (C_V - E_V) - 2E_V & = & 0, \\
(C_P + 2 E_V) - (C_P - E_V) - 3 E_V & = & 0.
\eea

For these modes, we obtain the following relations
\bea
|E_V|^2 &=& |A(\phi K^0)|^2, \\
|C_V|^2 &=& |A(\rho^0 K^0)|^2 + |A(\omega K^0)|^2 - |A(\phi K^0)|^2, \\
|C_P|^2 &=& 4|A(K^{*0}\eta^{\prime})|^2 + |A(K^{*0}\eta)|^2 - 2|A(\phi K^0)|^2, \\
\cos\delta_{C_V E_V} &=& \frac{1}{2|C_V||E_V|}
\left[ |A(\rho^0 K^0)|^2 - |A(\omega K^0)|^2 \right], \\
\cos\delta_{C_P E_V} &=& \frac{1}{2|C_P||E_V|}
\left[ |A(K^{*0}\eta)|^2 - 2|A(K^{*0}\eta^{\prime})|^2 - |A(\phi K^0)|^2
 \right].
\eea

These relations all suffer from the presence of a $K^0$ in at least one of
their amplitudes, and contamination by the corresponding mode with $\bar K^0$
makes them unusable.  However, the fact that with $E_P = - E_V$ we also
have amplitudes for the observable processes $D^0 \to (\rho^- K^+, K^{*0} \eta,
K^{*0} \eta')$, all of which involve $E_V$ and amplitudes which have been
previously specified, should allow the resolution of the last remaining
discrete ambiguity except in singular cases.

An analysis of SU(3) breaking based on the method of Ref.\ \cite{GRU} may be
able to provide direct information on relative strong phases in
Cabibbo-favored and doubly-Cabibbo-suppressed $D \to PV$ decays.  One needs
information on direct-channel resonances with $J^P = 0^-$, which is the
only channel which can decay to the $J = 0$ $PV$ state.  A candidate for
such a state around 1830 MeV (i.e., not far from the $D$ mass) has been
reported in the $K \phi$ channel \cite{Arm} but needs confirmation.

\section{Experimental prospects}

At present, the following WS modes are quoted by the Particle
Data Group: \cite{Groom:2000in}
\bea
\b(D^0 \to K^+\pi^-) &=& (1.46 \pm 0.30) \times 10^{-4}, \nonumber \\
\b(D^+ \to K^{*0}\pi^+) &=& (3.6 \pm 1.6) \times 10^{-4}, \nonumber \\
\b(D^+ \to \rho^0K^+) &=& (2.5 \pm 1.2) \times 10^{-4}, \nonumber \\
\b(D^+ \to \phi K^+) &<& 1.3 \times 10^{-4}.
\qquad ({\rm CL}=90\%) \nonumber
\eea
If one assumes that the amplitude $T$ is dominant in $PP$ modes, from the
branching ratio of $D^0 \to K^+\pi^-$ one would infer $\b(D^+ \to K^+\pi^0)
\simeq 1.8\times10^{-4}$ and $\b(D^+ \to K^+\eta) \simeq 1.2\times10^{-4}$.  A
substantial deviation from these expected values would indicate the importance
of $E$ and/or $A$ contributions.

Since the peak cross section for $e^+e^- \to \psi(3770) \to D {\bar D}$ is
about 10 nb and the foreseen integrated luminosity for a charm factory
operating at this energy is about 3 fb${}^{-1}$, one expects to collect
$3\times10^7$ $D{\bar D}$ pairs, giving about 15 million $D^0\,({\bar D}^0)$
and 15 million $D^+\,(D^-)$.  With branching ratios of $O(10^{-4})$ for the WS
decays, we would have $\sim 3000$ events for each type.  The $D^0$ decays must
be flavor-tagged through study of the flavor of the opposite-side neutral $D$.

Tagging via the chain $D^{*+} \to \pi^+ D^0$ is possible if one
operates at higher c.m. energy.  Indeed, it is estimated that in CLEO II.V
with 6 fb$^{-1}$ on the $\Upsilon(4S)$ and 3 fb$^{-1}$ in the continuum below
the $\Upsilon(4S)$, 34 million charmed mesons were produced \cite{DMA}.
BaBar and Belle should be able to accumulate an even larger sample.

In the analysis of $D \to PV$  decays, one needs to analyze the branching
ratios and resonant channel fractions of the set of 3-body final states listed
in Table \ref{tab:3bodymodes}.  Examples of recent progress in studying these
states are noted in Refs.\ \cite{Brandenburg:2001ze,Palano,FOCUS01,FOCUS99}.

\section{Conclusions}

As we have seen, doubly-Cabibbo-suppressed (``wrong-sign,'' or WS) decays with
a final neutral $K$ meson in general suffer from overwhelming backgrounds of
Cabibbo-favored (``right-sign,'' or RS) decays.
It is thus preferable to extract information from decay modes with charged $K$
mesons in the final states.  We have shown that the amplitudes for the $D^+$
decay modes ${K^+\pi^0,K^+\eta,K^+\eta'}$ form a triangle in the complex plane.
These charged $D$ decays provide a good place to study the amplitudes $|T|$,
$|A|$ and the relative strong phase $\cos\delta_{TA}$.  It will be interesting
to see whether in the case of WS $D$ decays one still observes $A$ and $E$ with
comparable amplitudes to $T$ and $C$ as in the RS decays \cite{Rosner:1999xd}.
It will also be useful to compare U-spin related RS and WS triangles to see
whether they are similar, from which one could learn final state interaction
patterns and U-spin breaking effects.

\begin{table}
{\caption{Summary of doubly-Cabibbo-suppressed 3-body modes required for
extracting amplitudes in $D \to PV$ decays.  All modes with a $K^0$ have $\bar
K^0$ backgrounds.  $D^+$ and $D_s^+$ modes with a $K^+$ are self-tagging.}
\label{tab:3bodymodes}}
\begin{center}
\begin{tabular}{ccc} \hline\hline
 & Final state & Branching ratio \\
\hline
$D^0$ &
\begin{tabular}{l}
$K^0\pi^+\pi^-$ \\ $K^+\pi^-\pi^0$ \\ $K^+\pi^-\eta$ \\ $K^+\pi^-\eta'$
\end{tabular} &
\begin{tabular}{l}
\\
$ (6.0\pm1.0)\times10^{-4}$ \cite{Brandenburg:2001ze} \\ 
$$ \\ $$
\end{tabular} \\
\hline
$D^+$ &
\begin{tabular}{l}
 $K^0\pi^+\pi^0$ \\
 $K^0\pi^+\eta$ \\
 $K^0\pi^+\eta'$ \\
 $K^+\pi^0\pi^0$ \\
 $K^+\pi^0\eta$ \\
 $K^+\pi^0\eta'$ \\
 $K^+\pi^-\pi^+$ \\
 $K^+K^+K^-$
\end{tabular} &
\begin{tabular}{l}
 \\
 \\
 \\
 \\
 \\
 \\
 $ (6.8\pm1.5)\times10^{-4}$ \cite{Groom:2000in}; see also \cite{FOCUS01} \\
 $ (1.41 \pm 0.27) \times 10^{-4}$ \cite{FOCUS99}
\end{tabular} \\
\hline
$D_s^+$ &
\begin{tabular}{l}
 $K^0K^0\pi^+$ \\
 $K^0 K^+ \pi^0$ \\
 $K^+ K^+ \pi^-$  
\end{tabular} &
\begin{tabular}{l}
 \\
 \\
\end{tabular} \\
\hline\hline
\end{tabular}
\end{center}
\end{table}

We also observed that without further assumptions, one could only form
quadrangle relations from the amplitudes for $D \to PV$ decays.  For example,
the four $D \to K^* \pi$ amplitudes form a quadrangle.  The relative phase
between the neutral $D$ amplitudes can be obtained by analyzing the $D^0 \to
K^+ \pi^- \pi^0$ Dalitz plot.  This fixes the quadrangle up to a two-fold
ambiguity corresponding to folding about the diagonal.
By further assuming $A_P=-A_V$, we can obtain three triangle relations and
determine $|T_V|$, $|T_P|$, $|A_V|$, $\cos\delta_{T_VA_V}$, and $\cos
\delta_{T_PA_V}$.  The two-fold quadrangle ambiguity can be resolved by
assuming $E_P = - E_V$ and measuring the rate for $D^0 \to K^+ \rho^-$.
Many cross-checks of the method are possible by measuring further WS
rates for three-body decays involving $\eta$ or $\eta'$ and by analyses of
interferences between right-sign and wrong-sign $K^* \pi$ contributions to
Dalitz plots.

\vspace{0.5cm}
{\it Acknowledgments:} We thank David Asner for stimulating discussions
and Michael Gronau, John Cumalat, and Harry Lipkin for helpful comments.
This work was supported in part by the United States
Department of Energy through Grants No.\ DE-FG02-90ER-40560 and W-31109-ENG-38.


\begin{thebibliography}{99}
\bibitem{HJLPC} H. J. Lipkin (private communication) has pointed out that
the linear combination of $\eta$ and $\eta'$ which enter these sum rules
is just the octet component, which should also be true for other choices
of octet-singlet mixing for the $\eta$ and $\eta'$.
\end{thebibliography}
\end{document}